\documentclass[aps,prl,showpacs,floatfix,twocolumn,amsmath,amssymb]{revtex4}

\usepackage{graphicx}
\usepackage{hyperref}

\begin{document}
\title{Theory of parametric amplification in superlattices}
\author{Timo Hyart}
\author{Alexey V. Shorokhov}
\author{Kirill N. Alekseev}
\altaffiliation{Also at: Department of Physics, Loughborough
University LE11 3TU, UK}%
\affiliation{Department of Physical
Sciences, P.O. Box 3000, FI-90014 University of Oulu, Finland}
\begin{abstract}
We consider a high-frequency response of electrons in a single
miniband of superlattice subject to dc and ac electric fields. We
show that Bragg reflections in miniband result in a parametric
resonance which is detectable using ac probe field. We establish
theoretical feasibility of phase-sensitive THz amplification at
the resonance. The parametric amplification does not require
operation in conditions of negative differential conductance. This
prevents a formation of destructive domains of high electric field
inside the superlattice.
\end{abstract}
\maketitle
Motion of an electronic wavepacket in a periodic lattice potential
with a period $a$ subject to a constant electric field $E_{dc}$ is
characterized by oscillations of its velocity with the Bloch
frequency $\omega_B=eaE_{dc}/\hbar$ \cite{bloch-zener}. Bloch
oscillations originate in Bragg reflections of the particle from
the Brillouin zone boundary. Among solid state structures,
artificial semiconductor superlattices (SLs) with a relatively
large period and narrow bands are most suitable for manifestation
of Bloch oscillations effects \cite{esaki70}. In the stationary
transport regime, Bloch oscillations causes static negative
differential conductivity (NDC) of SL if $\omega_B\tau>1$
($\tau\simeq 100$ fs is a characteristic scattering time)
\cite{esaki70}. For $\omega_B\tau>1$ and homogeneous distribution
of electric field inside SL, it can potentially provide a strong
gain for THz frequencies \cite{kss}. However, in conditions of
static NDC the same Bragg reflections, which give rise to Bloch
oscillations, do excite a soft dielectric relaxation mode
resulting in a formation of domains of high field inside SL
\cite{buettiker77}. The electric domains destroy Bloch gain in a
long SL. Therefore an utilization of Bloch gain is a difficult
problem \cite{wacker02}.
\par
Using simple semiclassical approach, let us  consider now an
influence of Bragg reflections on dynamics of an electron subject
to a strong ac (pump) field $E_p(t)=E_0\cos\omega t$. Combining
the acceleration theorem for the electron momentum along the SL
axis, $\dot{p}=eE_p(t)$, and the tight-binding energy-momentum
dispersion for a single miniband of the width $\Delta$,
$\varepsilon(p)=-(\Delta/2)\cos(pa/\hbar)$, we arrive to the
expression $\varepsilon(t)=\sum_{k=0}^{\infty} C_{2k}\cos(2k\omega
t)$, where $C_{2k}=-\Delta J_{2k}(eaE_0/\hbar\omega)$ for $k>0$
($J_n(x)$ are the Bessel functions). It shows that the electron
energy within the miniband varies with frequencies which are some
even harmonics of $\omega$: $\omega_\varepsilon^{even}=s\omega$
($s=2,4,6\ldots$). If a bias $E_{dc}$ is also included to the pump
field, $\varepsilon(t)$ oscillates with two combinations of
frequencies $\omega_B\pm\omega_\varepsilon^{even}$ and
$\omega_B\pm\omega_\varepsilon^{odd}$, where
$\omega_\varepsilon^{odd}=s\omega$ with $s=1,3,5\ldots$. However,
in the presence of collisions the oscillations with Bloch
frequency decay, whereas energy oscillations with the frequencies
imposed by ac field survive. The effective electron mass in the
nonparabolic miniband also varies periodically with the frequency
of energy oscillations. Now let us suppose that additionally a
weak ac field $E_{pr}=E_1\cos(\omega_1 t+\phi)$ is also applied.
The frequency of this probe field $\omega_1$ is fixed by an
external circuit (resonant cavity). Since electron transport in
the band depends on an instant value of the effective electron
mass, one should expect the parametric resonance for
$\omega_\epsilon^{(s)}=l\omega_1$ ($l$ is an integer and
$\omega^{(s)}_\varepsilon$ stands for either
$\omega_\varepsilon^{even}$ or $\omega_\varepsilon^{odd}$). The
most strong parametric resonance occurs when $l=2$, that is for
$\omega^{(s)}_\varepsilon/2=\omega_1$ \cite{migulin}. As in other
parametric devices \cite{migulin}, the parametric resonance due to
Bragg reflections can result in a regenerative amplification of
the probe field. However, currents at harmonics of the pump ac
field are generated in SL due to strong nonparabolicity of its
miniband \cite{esaki71}. If the parametric amplification arises at
the same frequencies as the frequencies of generated harmonics,
the effect of harmonics  blurs out the weaker ($\propto E_1$)
effect of small-signal gain. This problem is well-known for the
parametric amplification in Josephson junctions, which also have
strong nonlinearity \cite{param-jj}.
\par
We are interested in manifestations of the parametric resonance
due to Bragg reflections in the presence of collisions, i.e. in
the miniband transport regime \cite{wacker02}. Here two main
questions arise: Can the parametric resonance provide a
high-frequency gain in the miniband transport regime? Is it
possible to avoid space-charge instability? Some of these problems
have been discussed earlier. In 1977 Pavlovich first used
Boltzmann transport approach to calculate the coefficient of
intraband absorption of a weak probe field ($\omega_1$) in  SL
subjected to a strong ac pump of commensurate frequency ($\omega$)
\cite{pavlovich}. He briefly mentioned a possibility of negative
absorption for some $\omega_1/\omega$. However, neither physical
origin of the effect nor its compatibility with conditions of
electric stability were addressed in this pioneer work. Further,
in a recent letter \cite{hyart06}, we presented numerical support
for a possibility of parametric amplification without formation of
electric domains in the miniband transport regime. Solving
numerically balance equations for SL \cite{ignatov76} we
demonstrated a feasibility of gain at even harmonics. In this
situation, we observed that gain can exist in the absence of NDC.
It guarantees electric stability for moderate concentrations of
electrons \cite{alekseev06}.
\par
\begin{figure}
\includegraphics[width=1\columnwidth,clip=]{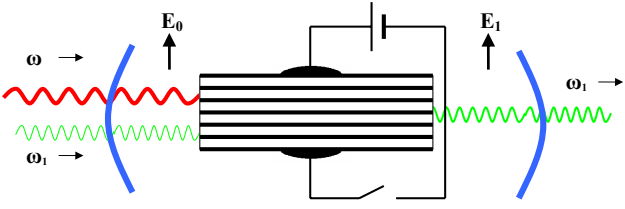}
\caption{\label{fig1}(color online) Two schemes of the parametric
amplification in superlattice without corruption from generated
harmonics. In the presence of ac pump (red online) of the
frequency $\omega$, parametric gain for a weak signal (green
online) of the frequency $\omega_1$ arises either at
$\omega_1=:\omega/2,3\omega/2,\ldots$ in biased SL ($E_{dc}\neq
0$) or at $\omega_1=: 2\omega,4\omega,\ldots$ in unbiased SL
($E_{dc}=0$).}
\end{figure}

In this paper, we analytically calculate gain of a weak
high-frequency (THz) probe field in SL miniband under the
conditions of parametric resonance,
$\omega^{(s)}_\varepsilon/2=\omega_1$, caused by the action of a
strong ac pump field. The physical origin of the parametric
resonance is  a periodic variation of effective electron masses in
miniband and, at high THz frequencies, also a variation of
specific quantum inductance. We prove that for a proper choice of
relative phase $\phi$ a power is always transferred from the pump
to the probe field. Furthermore, we show that the same pump field
also modifies free carrier absorption in SL. We find that the gain
caused by the parametric resonance can sufficiently overcome the
modified free carrier absorption and simultaneously remain
unaffected by the generated harmonics of the pump only in two
distinct cases: For amplification at half-harmonics in biased SL
and for amplification at even harmonics in unbiased SL
(Fig.~\ref{fig1}). In both these cases we predict a significant
amplification at room temperature in the absence of NDC.
\par
Within the semiclassical approach \cite{wacker02} we first solved
Boltzmann transport equation for a single miniband and bichromatic
field $E_p(t)+E_{pr}(t)$ with commensurate frequencies. Then we
calculated the phase-dependent absorption of the probe field,
which is defined as
\begin{equation}
\label{A-def} %
A=\langle V(t) \cos(\omega_1 t+ \phi)\rangle_t,
\end{equation}
where $V(t)=\overline{V}(t)/V_{p}$ is the electron velocity
$\partial\varepsilon(p)/\partial p$ averaged over a distribution
function satisfying the Boltzmann equation and
$\langle\ldots\rangle_t$ means averaging over a time period which
is common for both pump ($\omega$) and probe ($\omega_1$) fields.
Gain corresponds to $A<0$. Note that through the paper the
averaged velocity $V$, averaged energy
$W=\overline{\varepsilon}/|\varepsilon_{eq}|$ and field strengths
$E_{0,1}$ and $E_{dc}$ are scaled to the Esaki-Tsu peak velocity
$V_p=(\Delta a/4\hbar)\mu(T)$, the equilibrium energy in absence
of fields $\varepsilon_{eq}=-(\Delta/2)\mu(T)$ \cite{wacker02} and
the critical field $E_c=\hbar/ea\tau$ \cite{esaki70},
respectively. The temperature factor is $\mu(T)=I_1(\Delta/2k_B
T)/I_0(\Delta/2k_B T)$ (here $I_{0,1}(x)$ are the modified Bessel
functions) \cite{allen90}. Absorption of a weak ($E_1\rightarrow
0$) probe field  in SL (Eq.~\ref{A-def}) is linear in $E_1$. It
can be naturally represented as the sum of phase-dependent
coherent and phase-independent incoherent components
$A=A_{coh}+A_{inc}$.
\par
Parametric effects in the absorption are described by its coherent
component. It has the form
\begin{equation}
\label{A_coh-general}
A_{coh}=-\left(\beta_1/4\right)
B\cos[2(\phi-\phi_{opt})],
\end{equation}
where the amplitude of coherent absorption $B>0$ and
$\beta_1=E_1/(\omega_1\tau)$. The coherent component always
provides gain if $|\phi-\phi_{opt}|<\pi/4$. Gain has maximum at an
optimal phase $\phi_{opt}$. Under the action of pump field, such
energy storage parameters of SL as the energy of electrons in
miniband $W$ and mesoscopic electric reactance, which is described
by the reactive current $I^{\sin}\propto V^{\sin}$, are
simultaneously harmonically modulated. The variables $B$ and
$\phi_{opt}$ can be represented in terms of the specific harmonics
of $W(t)$ and out-of-phase component of electron velocity
$V^{\sin}(t)$ as
\begin{equation}
\label{B_opt-phi_opt}
B=\left[ B_{lf}^2+B_{hf}^2\right]^{1/2},\quad
\tan(2\phi_{opt})=-B_{hf}/B_{lf},
\end{equation}
\begin{eqnarray}
\label{B_1-B_2}
B_{lf}&=&2W^{\sin}_s(\omega_B),\nonumber\\
B_{hf}&=&
W^{\cos}_s(\omega_B+\omega_1)-2W^{\cos}_s(\omega_B)+W^{\cos}_s(\omega_B-\omega_1)\nonumber\\
   & & \mbox{}
   -\left[V^{\sin}_s(\omega_B+\omega_1)-V^{\sin}_s(\omega_B-\omega_1)\right],
\end{eqnarray}
where the index $s$ is the same as involved in the condition of
parametric resonance and the Fourier components of the quantum
reactive parameters are given by
\begin{eqnarray}
\label{explicit}
W^{\cos}_k&=&-\sum_l J_l(\beta)\left[J_{l-k} (\beta) +
J_{l+k} (\beta)\right] K(\omega_B+l\omega),\nonumber\\
V^{\sin}_k&=&\sum_{l}
J_l(\beta)\left[J_{l+k}(\beta)-J_{l-k}(\beta)
\right] K(\omega_B+l\omega),\\
W^{\sin}_k&=&-\sum_l J_l(\beta)\left[J_{l-k}(\beta)
-J_{l+k}(\beta)\right] V^{ET}(\omega_B+l\omega).\nonumber
\end{eqnarray}
In Eqs.~\ref{explicit} $\beta=E_0/(\omega\tau)$, the Esaki-Tsu
drift velocity
$V^{ET}(\omega_B)=\frac{\omega_B\tau}{1+(\omega_B\tau)^2}$
\cite{esaki70,wacker02} and Esaki-Tsu energy
$K(\omega_B)=\frac{1}{1+(\omega_B\tau)^2}$
\cite{ignatov76,wacker02} determine the dependence of
$W^{\cos}_k(\omega_B)$, $W^{\sin}_k(\omega_B)$ and
$V^{\sin}_k(\omega_B)$ on the dc bias $E_{dc}$. It worth to notice
that instead of harmonics of energy we alternatively can consider
harmonics of effective electron mass because
$m^{-1}(\overline{\varepsilon})\propto W$.
\par
In the low frequency range  $\omega\tau,\omega_1\tau\ll 1$, we
found that $B_{hf}\rightarrow 0$ and therefore $B=B_{lf}$, while
for THz frequencies ($\omega\tau\gtrsim 1$) both terms  $B_{lf}$
and $B_{hf}$ contribute to $B$. The behavior of the absorbtion
amplitude $B$ at THz frequencies has two peculiarities. First,
influence of the out-of-phase component of electron velocity at
the pump frequency and its harmonics also becomes important. As
follows from Eq.~(\ref{explicit}), it describes inductive response
of inertial miniband electrons to ac field in the limit
$\omega\tau\gg1$: $V^{\sin}_1=E_0/\omega\tau L$,
$L^{-1}=2J_0(\beta)J_1(\beta)\beta^{-1}K(\omega_B)$
\cite{ghosh-apl}. Second, interaction of miniband electrons with
THz fields has quantum nature \cite{wacker02}. Therefore, even a
very weak probe field produces a back action on the SL reactive
parameters. This is indicated by an appearance of virtual
processes of absorption and emission of one quantum of the probe
field ($\pm\hbar\omega_1$) in the expression for $B_{hf}$
(Eq.~(\ref{B_1-B_2})). In particular, $B_{hf}$ is determined by
the difference between changes in electron energy at absorption
$W(\omega_B+\omega_1)-W(\omega_B)$ and emission
$W(\omega_B)-W(\omega_B-\omega_1)$. The asymmetry in the
elementary acts of emission and absorption is caused by
scattering. It resembles corresponding asymmetry revealed in the
quantum description of THz Bloch gain in dc biased SLs
\cite{willenberg}.
\par
We turn now to the analysis of the incoherent component of
absorption $A_{inc}$, which is independent on both the ratio
$\omega_1/\omega$ and phase difference $\phi$. It can be
represented as
\begin{equation}
\label{A_inc}
A_{inc}=\frac{\beta_1}{2} \left[
V_{dc}(\omega_B+\omega_1)-V_{dc}(\omega_B-\omega_1)\right],
\end{equation}
where $V_{dc}=\langle V\rangle_t$ is the drift velocity induced in
SL by the pump field alone. It is determined by the well-known
formula \cite{unter_and_refs}
\begin{equation}
\label{V_dc}
V_{dc}(\omega_B)=\sum_l J_l^2(\beta)V^{ET}(\omega_B+l\omega).
\end{equation}
$A_{inc}$ describes the free carrier absorption modified by the
pump. Naturally, $A_{inc}$ becomes the usual free carrier
absorption $A_{inc}\propto(1+\omega_1^2\tau^2)^{-1}$ in the
absence of pump field ($E_0=E_{dc}=0$). Remarkably, as follows
from Eq.~(\ref{A_inc}), the pump  could suppress the free carrier
absorption (if $V_{dc}(\omega_B+\omega_1)\approx
V_{dc}(\omega_B-\omega_1)$) or even make its value negative (if
$V_{dc}(\omega_B-\omega_1)>V_{dc}(\omega_B+\omega_1)$).
\par
On the other hand, it is easy to see that in the quasistatic
limit, $\omega_1\tau\ll 1$, the finite difference in
Eq.~(\ref{A_inc}) goes to the derivative $\partial V_{dc}/\partial
E_{dc}$, which determines the slope of dependence of $V_{dc}$ on
dc bias at the working point $E_{dc}$. The sign of this derivative
controls electric stability against spatial perturbations of
charge density \cite{kroemer_eprint,alekseev06}: For negative
slope $\partial V_{dc}/\partial E_{dc}<0$ destructive space-charge
instability arises inside SL. In contrast, $\partial
V_{dc}/\partial E_{dc}>0$ is the necessary condition for absence
of the electric domains in moderately doped SLs \cite{alekseev06}.
\begin{figure}
\includegraphics[width=1\columnwidth,clip=]{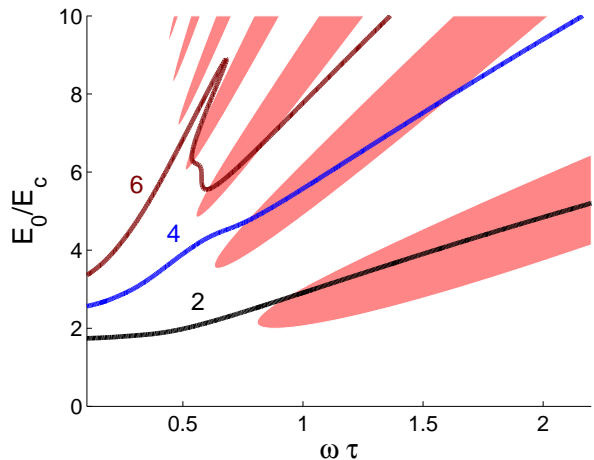}
\caption{\label{fig2}(color online) Amplification at even
harmonics in unbiased superlattice for $\phi=\phi_{opt}$. Regions
above the marked curves correspond to gain at
$\omega_1=:2\omega,4\omega,6\omega$. Dark (red online) areas
correspond to electric instability.}
\end{figure}
\par
For general case $\omega_1\tau\gtrsim 1$ our numerical analysis
showed that the sign of finite difference (\ref{A_inc}) is almost
always same as the sign of the derivative $\partial
V_{dc}/\partial E_{dc}$, if SL is unbiased ($E_{dc}=0$) or only
weakly biased. Therefore $A_{inc}>0$ guarantees electric
stability. The total absorption $A=A_{coh}+A_{inc}$ still can be
negative in conditions of electric stability if
$|\phi-\phi_{opt}|<\pi/4$ and $\mid A_{coh}\mid>A_{inc}$.
\begin{figure}
\includegraphics[width=1\columnwidth,clip=]{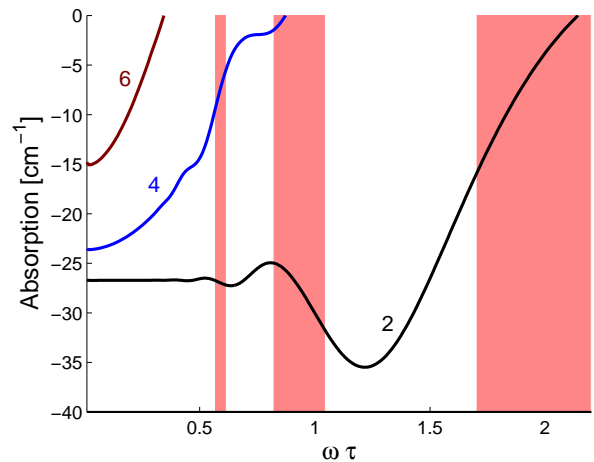}
\caption{\label{fig3}(color online) Magnitude of negative
absorption at even harmonics (marked curves) as a function of the
pump frequency $\omega$ for the fixed pump amplitude $E_0=5.1$ and
$\phi=\phi_{opt}$. Dark (red online) segments indicate intervals
of NDC.}
\end{figure}
In the case of unbiased SL (Fig.~\ref{fig1}), such situation is
illustrated in Figs~\ref{fig2},\ref{fig3}. Fig.~\ref{fig2} shows
the regions of negative absorption ($A<0$) at even harmonics
together with the regions of NDC ($\partial V_{dc}/\partial
E_{dc}<0$) in $\omega E_0$ plane. Here the phase is chosen to be
optimal (Eq.~\ref{B_opt-phi_opt}). The values of $E_0$ and
$\omega$ resulting in electric instability (red areas in
Fig.~\ref{fig2}) are close to the lines of Bessel roots
$J_0(\beta)=0$. It can be explained noticing that for
$E_{dc}\rightarrow 0$ transition to NDC is accompanied by absolute
negative conductivity (ANC) \cite{alekseev06}. However, as can be
derived from Eq.~\ref{V_dc} in the limit $E_{dc}\rightarrow 0$,
ANC arises only for $J_0(\beta)\simeq 0$ \cite{ignatov76}.
Importantly, the regions of gain and areas of instability overlap
only in limited ranges of the pump amplitudes and frequencies.
Moreover, the magnitude of domainless gain is significant even at
room temperature (Fig.~\ref{fig3}). To estimate gain $\alpha$ in
units cm$^{-1}$ \cite{willenberg} we used the formula
$\alpha=\alpha_0\times (A/E_1)$ with $\alpha_0=8\pi e N V_p/(E_c
n_r c)$ and the following typical semiconductor SL parameters:
$a=6$ nm, $\Delta=60$ meV, electron density $N={10}^{16}$
cm$^{-3}$, $\tau=200$ fs, refractive index $n_r=\sqrt{13}$ (GaAs)
and $T=300$ K.
\par
\begin{figure}
\includegraphics[width=0.9\columnwidth,clip=]{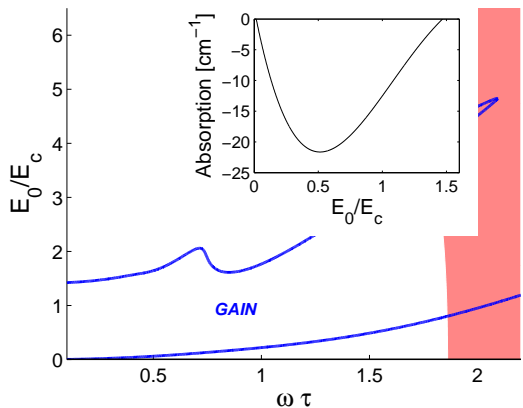}
\caption{\label{fig4}(color online) Amplification with a low
threshold at $\omega_1=\omega/2$ in biased superlattice for
$E_{dc}=1$ and $\phi=\phi_{opt}$. Marked region between curves
corresponds to gain, while dark (red online) area corresponds to
electric instability. Inset: Magnitude of negative absorption as
function of the pump amplitude $E_0$ for $\omega\tau=0.25$.}
\end{figure}

Even harmonics of the pump satisfy the parametric resonance
condition $\omega_\varepsilon^{even}/2=\omega_1$. For unbiased
case only this scheme provides amplification which is unaffected
by generated harmonics. On the other hand, for $E_{dc}\neq 0$
subharmonics of the pump ($\omega_1=:\omega/2,3\omega/2,\ldots$)
satisfy another parametric resonance condition
$\omega_\varepsilon^{odd}/2=\omega_1$. We found that regions of
gain at different half-harmonics  and areas of electric
instability (NDC) have no overlapping for many values of $E_0$ and
$\omega\tau$. Fig.~\ref{fig4} illustrates this for amplification
at $\omega_1=\omega/2$ and $E_{dc}=1$.
Here threshold is very low while gain is still significant even at
$E_0\leq 0.5$ (Fig.~\ref{fig4}, inset). We explain it analyzing
the behavior of both $A_{coh}$ and $A_{inc}$ for small $E_0$.
First, for $\omega_1/\omega=1/2$ relatively large first harmonics
($s=1$) of the quantum reactive parameters contribute to
$A_{coh}<0$ (Eq.~\ref{B_1-B_2}). Second, the tangent to the curve
describing a dependence of $V_{dc}$ on $E_{dc}$ (Eq.~\ref{V_dc})
has a small positive slope at the working point $E_{dc}=1$.
Following Eq.~\ref{A_inc} it results in a rather small
$A_{inc}>0$. Therefore the total gain $A<0$ is not small.
\par
In this Letter, we focused on the phase-sensitive degenerate
parametric amplification of THz fields in SLs. Our theory can be
directly extended to describe nondegenerate phase-insensitive
amplification. Here, at least for the case of unbiased SL, regions
of NDC in $\omega E_0$ plane are still located only near Bessel
roots lines (\textit{cf}. Fig.~\ref{fig2}). Therefore, by a proper
choice of amplitude and frequency of the pump it is also possible
to reach electrically stable amplification of weak signal
($\omega_1$) and idler ($\omega_2$) fields satisfying the
parametric resonance condition
$\omega_\varepsilon^{even}=\omega_1+\omega_2$.
\par
The parametric effects in a nonparabolic energy band should exist
not only in semiconductor SLs but also in other artificial
periodic structures, including periodic waveguide arrays
\cite{wavegu_sl} and microcavity SLs \cite{photcr_sl} for light,
phononic microcavity arrays \cite{phonon_sl}, carbon nanotube SLs
in perpendicular electric field \cite{nanotube_sl}, and
dissipative optical lattices for ultracold atoms \cite{ott04}.
These SLs were specially suggested and designed to manifest
effects of Bloch oscillations
\cite{photcr_sl,phonon_sl,nanotube_sl,ott04} or ac field
\cite{wavegu_sl,nanotube_sl} in a single band  and therefore
potentially can be used to observe the parametric amplification.
\par
In summary, we described physical mechanisms for the parametric
resonance and resulting high-frequency amplification in an energy
band. The parametric amplification of a weak signal is possible
without negative differential conductance. Parametric effects due
to Bragg reflections in ac-driven lattices are no less important
than manifestations of Bloch oscillations in the case of a pure dc
bias.
\par
This work was supported by Academy of Finland, Emil Aaltonen
Foundation, and AQDJJ Programme of ESF.


\begin{thebibliography}{99}

\bibitem{bloch-zener}
C. Zener, Proc. R. Soc. London Ser. A \textbf{145}, 523 (1934).

\bibitem{esaki70}
L. Esaki and R. Tsu, IBM J. Res. Dev. \textbf{14}, 61 (1970).

\bibitem{kss}
S. A. Ktitorov, G. S. Simin, and V. Ya. Sindalovskii, Sov. Phys.
Solid State \textbf{13}, 1872 (1972).

\bibitem{buettiker77}
M. B\"{u}ttiker and H. Thomas, Phys. Rev. Lett. \textbf{38}, 78
(1977).

\bibitem{wacker02}
A. Wacker, Phys. Rep. \textbf{357}, 1 (2002).

\bibitem{migulin}
V. V. Migulin, \textit{Basic Theory of Oscillations} (Mir, Moscow,
1983).

\bibitem{esaki71}
L. Esaki and R. Tsu, Appl. Phys. Lett. \textbf{19}, 246 (1971).

\bibitem{param-jj}
M. J. Feldman, P. T. Parrish, and R. Y. Chiao, J. Appl. Phys.
\textbf{46}, 4031 (1975); R. Movshovich, B. Yurke, A. D. Smith,
and A. H. Silver, Phys. Rev. Lett. \textbf{67}, 1411 (1991).

\bibitem{pavlovich}
V. V. Pavlovich, Fiz. Tverd. Tela (Leningrad) \textbf{19}, 97
(1977) [Sov. Phys. Solid State \textbf{19}, 54 (1977)].

\bibitem{hyart06}
T. Hyart, N. V. Alexeeva, A. Lepp\"{a}nen, and K. N. Alekseev,
Appl. Phys. Lett. \textbf{89}, 132105 (2006).

\bibitem{ignatov76}
A. A. Ignatov and Yu. A. Romanov, Phys. Stat. Sol. B {\bf 73}, 327
(1976).

\bibitem{alekseev06}
K. N. Alekseev \textit{et al}., Europhys. Lett. \textbf{73}, 934
(2006).

\bibitem{allen90}
G. Brozak \textit{et al}., Phys. Rev. Lett. \textbf{64}, 3163
(1990).

\bibitem{ghosh-apl}
A. W. Ghosh, M. C. Wanke, S. J. Allen, and J. W. Wilkins, Appl.
Phys. Lett. \textbf{74}, 2164 (1999).

\bibitem{willenberg}
H. Willenberg, G. H. D\"{o}hler, and J. Faist, Phys. Rev. B
\textbf{67}, 085315 (2003).

\bibitem{unter_and_refs}
V. V. Pavlovich and E. M. Epshtein, Sov. Phys. Semicond.
\textbf{10}, 1196 (1976); K. Unterrainer \textit{et al}., Phys.
Rev. Lett. \textbf{76}, 2973 (1996).

\bibitem{kroemer_eprint} H. Kroemer, cond-mat/0009311 .

\bibitem{wavegu_sl}
S. Longhi \textit{et al}., Phys. Rev. Lett.  \textbf{96}, 243901
(2006).

\bibitem{photcr_sl}
V. Agarwal \textit{et al}., Phys. Rev. Lett. \textbf{92}, 097401
(2004).

\bibitem{phonon_sl}
N. D. L. Kimura, A. Fainstein, and B. Jusserand, Phys. Rev. B
\textbf{71}, 041305 (2005).


\bibitem{nanotube_sl}
O. V. Kibis, D. G. W. Parfitt, and M. E. Portnoi, Phys. Rev. B
\textbf{71}, 035411 (2005).

\bibitem{ott04}
H. Ott \textit{et al}., Phys. Rev. Lett. \textbf{92}, 160601
(2004).

\end{thebibliography}
\end{document}